\documentclass[pre,amssymb,aps,twocolumn]{revtex4}
\usepackage{array}
\usepackage{epsfig}

\def\be{\begin{equation}}
\def\ee{\end{equation}}
\def\br{\begin{array}}
\def\er{\end{array}}
\def\ba{\begin{eqnarray}}
\def\ea{\end{eqnarray}}
\def\bfg{\begin{figure}}
\def\efg{\end{figure}}

\def\vr{\varrho}

\def\vrSB{\varrho_{SB}}

\def\sp8{\sqrt{\pi\over 8}}

\def\avrho{\langle\vr\rangle}

\begin{document}
\title{Shearing of loose granular materials: A statistical 
mesoscopic model}
\author{J\'anos T\"or\"ok$^{(1)}$, Supriya Krishnamurthy$^{(2,*)}$,
J\'anos Kert\'esz$^{(1)}$ and St\'ephane Roux$^{(3)}$}
\affiliation{(1): Department of Theoretical Physics,
Institute of Physics, Budapest University of Technology and Economics,
8 Budafoki \'ut, H-1111 Budapest, Hungary \\
(2): Department of Theoretical Physics, University of Oxford,
1 Keble Road, OX1 3NP, UK\\
(3): Surface du Verre et Interfaces, UMR CNRS/Saint-Gobain,
39 Quai Lucien Lefranc, 93303 Aubervilliers Cedex, France}
\thanks{present address: Santa Fe Institute, 1399 Hyde Park Road, 
Santa Fe NM 87501}

\begin{abstract}
A two-dimensional lattice model for the formation and evolution of
shear bands in granular media is proposed. Each lattice site is
assigned a random variable which reflects the local density. At every
time step, the strain is localized along a single shear-band which is
a spanning path on the lattice chosen through an extremum condition.
The dynamics consists of randomly changing the `density' of the sites
only along the shear band, and then repeating the procedure of
locating the extremal path and changing it. Starting from an initially
uncorrelated density field, it is found that this dynamics leads to a
slow compaction along with a non-trivial patterning of the system,
with high density regions forming which shelter long-lived low-density
valleys. Further, as a result of these large density fluctuations, the
shear band which was initially equally likely to be found anywhere on
the lattice, gets progressively trapped for longer and longer periods
of time. This state is however meta-stable, and the system continues
to evolve slowly in a manner reminiscent of glassy dynamics. Several
quantities have been studied numerically which support this picture
and elucidate the unusual system-size effects at play.
\end{abstract}

\date{\today}
\maketitle

\section{Introduction}

Modeling the rheology of granular media using continuum solid
mechanics, has reached a high degree of sophistication in terms of
constitutive equations \cite{Wood}. Whatever the complexity of load
paths being studied, an accurate account of the experimental stress
strain relationship can now be achieved provided enough parameters or
internal variables are included in the constitutive laws. However,
such approaches are descriptive and leave unanswered questions
pertaining to the scale of grain sizes.

In parallel to such phenomenological descriptive theories, a lot of
effort has been spent in recent years in developing powerful computer
models able to simulate granular systems at the individual grain
level. Molecular dynamics approaches \cite{LudingMD}, or other
techniques such as ``contact dynamics''
\cite{Radjai_Brendel,Radjaiforce}, now offer the possibility of
dealing with several thousand particles, and provide extremely
realistic pictures of the detailed micro-mechanics. 

Such numerical techniques can be used to accurately investigate
displacement fields, resolved both spatially and
temporally\cite{Radj}. The latter reveal an intriguing feature: namely
that even in the most simple tests, such as a simple steady shear
imposed over large strains, the local displacement appears very
unsteady, with short quiescent periods where the displacement field is
spatially smooth, separated by sudden changes where the configuration
of grains reaches a local instability and undergoes a rapid
reorganization through significant displacements at the grain level.
This temporal variability manifests itself in giant stress
fluctuations observed experimentally when particles and walls are
stiff, and when the high frequency part of the stress signal is not
filtered out\cite{Behr}. Numerical simulations indicate \cite{More}
that instantaneous strain fields consist essentially in localized
strains occurring along one or a few shear bands. However, as the
strain increases (over the moderate range accessible in the
simulation), there seems to be little or no correlation between
successive shear bands, so that the time average of the displacement
field erases these discontinuities and produces smooth strain fields. 

Such fluctuations are obviously ignored in continuum modeling. And
indeed it may appear that the identification of these instabilities is
relevant only for discussing fine details of microscopic and transient
features. However, their relevance can be judged only at a mesoscopic
level of modeling, since the microscopic numerical techniques are far
from being able to reach the relevant time scales. In fact, in the
following we will argue that these instabilities may have a
significant impact, both on large scale heterogeneities of the medium
itself, and on a systematic slow time evolution of the macroscopic
friction angle. A short account of some of our results has appeared in
a previous publication \cite{TKKRLett}.

The paper is organized as follows: in Section II, we will recall some
features observed experimentally or numerically that we consider
essential, and, in Section III, we progressively introduce the rules
of a model whose aim is to describe some statistical aspects of
shearing of loose granular media over large strains. In Section IV, we
present in details the different quantities studied numerically for
this model. We conclude in section V with a summary of our results and
a discussion on possible experimental checks.

%%%%%%%%%%%%%%%%%%%%%%%%%%%%%%%%%%%%%%%%%%%%%%%%%%%%%%
\section{The shear process in loose granular material}
%%%%%%%%%%%%%%%%%%%%%%%%%%%%%%%%%%%%%%%%%%%%%%%%%%%%%%

We will address here the question of the behavior of granular media
subjected to a simple shear for large strains. We restrict ourself to
the simplest granular medium one may consider, namely rigid
(undeformable) grains with Coulomb friction. This refers
experimentally to dry sand subjected to a low confining pressure. We
are concerned here with large strains, and thus in order to avoid the
problem of boundary conditions which would limit the maximum strain,
we consider an annular shear cell. To simplify the problem further, we
consider only the case where the problem is invariant along the shear
direction. As shown in Figure \ref{Fig_descr}, the displacement is a
single function of the coordinate of a radial cross-section $(x,y)$,
and constant along the orthoradial direction $z$ (traditionally this
situation is termed ``anti-plane''). Moreover, we are interested only
in the quasistatic regime, i.e., time as such is irrelevant, and only
the total strain matters. Thus, in what follows, what is referred to
as ``time'' is to be seen here as a practical means of parameterizing
the total strain being imposed on the medium. 

\bfg
\centerline{\epsfig{file=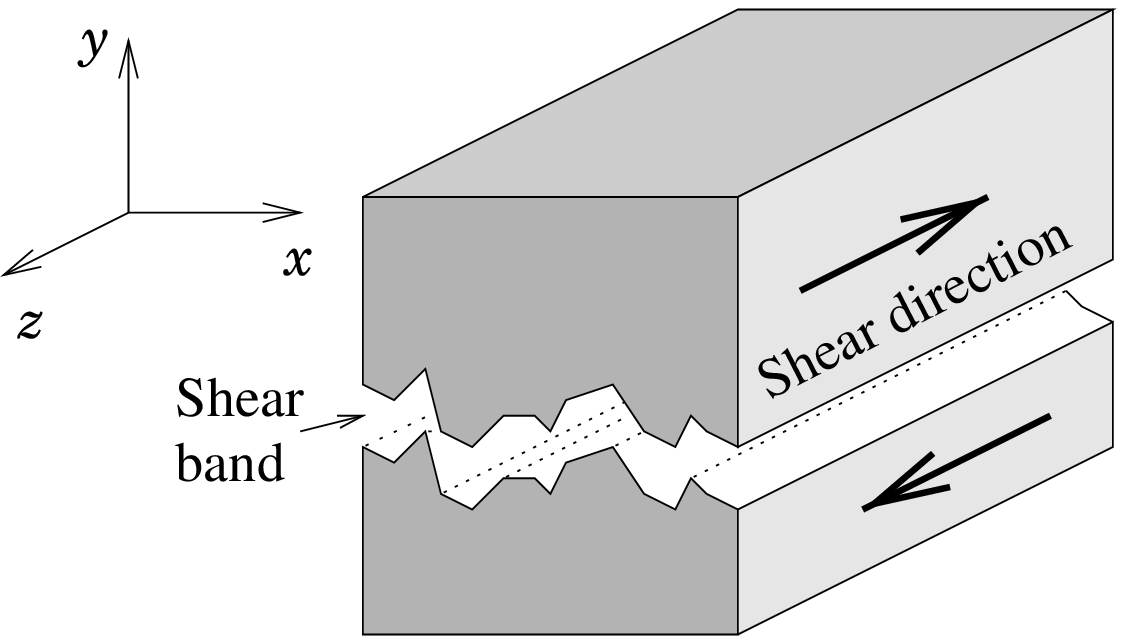,width=7truecm}}
\medskip
\caption{\label{Fig_descr}\small
Schematic picture of the shear process. The shear band is parallel to
the shear direction $z$ due to the periodic boundary conditions in
this direction. We sum up along this direction to get a two
dimensional sample in the $xy$ plane.}
\efg

One of the important observations of soil mechanics concerning such
media is the concept of a critical state \cite{Critstate1,Critstate2}.
Depending on the preparation of the sample, the behavior under shear
may differ considerably. For loose sand, (low density), the deviatoric
stress to be applied increases with the total shear strain and
simultaneously, a densification is observed
\cite{CompactExp,LudingModel}. However, as the shear strain increases,
the density and shear stress seem to reach a plateau independent of
the initial density. This state is called the ``critical state''. On
the contrary, if the initial density is large, a single shear band
forms, while the rest of the medium remains frozen \cite{widthsh}.
The formation of the shear band is preceded by a volume expansion of
the medium \cite{dil}, but after the band is formed, all further
properties remain quasi constant. A detailed experimental
investigation has revealed \cite{Desrues} that inside the shear band,
the density tends to approach the critical state density. This concept
of the critical state has received considerable experimental evidence
over the years, and is implemented in a number of continuum
constitutive laws. Experiments however mostly deal with a rather
moderate total strain well below unity. 

A simple picture which is consistent with the critical state concept
is that both the friction and the dilation angle increase with the
density, and that the critical state is the density for which the
dilation angle is zero (no change in volume under shear). Retaining
the density as the only internal variable is an approximation. Other
characteristics of the texture of the medium such as the fabric tensor
(which has information about the orientation of contact normals),
certainly play a significant role. For the purpose of simplicity, we
will in the following only retain one single scalar internal variable
governing the friction angle. It could be either simply the density or
a combination of density and texture. Nevertheless, in all cases we
will refer to this internal variable as ``density'', irrespective of
its precise meaning.

As mentioned above, numerical simulations seem to reveal \cite{More}
evidence for the existence of instantaneous shear bands even in loose
granular media. On the other hand, in experiments the strain appears
to be homogeneous and not localized. The resolution of this apparent
paradox is that the shear bands change rapidly, and may visit the
entire medium in the process. Thus during an increment of shear, which
can be observed experimentally, only a time average over many such
shear bands is seen. In our modeling, we introduce a basic time scale
for each elementary procedure. This time step is then clearly much
shorter than most experimentally accessible time scales. However, our
model attempts to achieve a qualitative rather than a precise
quantitative mapping. One of the main features of the model is to show
that these two apparently unrelated facts: the existence of
instantaneous shear bands at early times and its localization at late
times, are actually related, with a slow transition between these two
limiting cases. This slow dynamics is reminiscent of slow ageing
properties encountered in glassy systems, and indeed, we will see that
a breakdown of ergodicity does appear in this model.

%%%%%%%%%%%%%%%%%%%%%%%%%%%%%%%%%%%%%%%
\section{The model}
%%%%%%%%%%%%%%%%%%%%%%%%%%%%%%%%%%%%%%%
\subsection{Motivation and definition}
%%%%%%%%%%%%%%%%%%%%%%%%%%%%%%%%%%%%%%%

At every instant the two-dimensional medium is characterized by a
single, scalar internal variable, the density $\vr(x,y)$. This
represents an average of the density along the orthoradial direction
$z$. From this density, we deduce a corresponding local friction
coefficient $\mu(x,y)$. The latter is assumed to be a single
monotonically increasing function of the density \cite{JanoPull}. For
simplicity, we may assume a linear relationship in the following
although this is inessential.

The strain is imposed on the shear cell through prescribed
displacements of the bottom and top planes. As particles are
considered rigid, (no elastic deformation), the shear cell can only
move if the shear force exceeds a threshold value proportional to the
normal pressure. This limit stress is given by the ``weakest internal
surface''. Indeed, in our anti-plane geometry, the shear strain will
localize on the surface (i.e. path in the $(x,y)$ plane) which will
fail first. The latter is assumed to be given by the following
algorithm. For each directed path $\cal P$ spanning the entire
cross-section along the $x$-axis, we compute the maximum shear force
it can support according to the local density. Assuming that the local
slope of the path is always small, this maximum force $F({\cal P})$ is
simply proportional to the sum of local friction coefficients, and
thus making use of the assumed linear variation of the friction
coefficient with the density, $F({\cal P})$ is proportional to the sum
of local densities, 
\be 
S({\cal P})=\sum_{(x,y)\in{\cal P}} \vr(x,y)\;,
\ee
where the sum runs over the sites along the path. Among all the
possible paths, the weakest ${\cal P}^*$ (for which $S({\cal
P}^*)=\min $) will fail first, and this fixes the value of the shear
force $F=F({\cal P}^*)$. In agreement with the previously mentioned
observation, at every basic time step, the shear strain is realized
along a single shear band. Away from this shear band, the strain rate
is zero, and thus the density is kept constant in time. However,
inside the shear band, there is a relative motion of grains, and thus
the density is susceptible to evolve.

The next step is now to determine how the density inside the shear
band evolves with time. Even though it is observed that large strains
are necessary in order for a system to reach its critical state, we
argue that at the microscopic level, the evolution of the medium {\it
cannot} depend on the total imposed strain. Thus the evolution rules
for the density within the shear band should be designed in such a way
that they do not depend on the past history, but only on the present
state (density field). As the density ought to contain the basic
information of the local characteristics, we propose that within the
shear band, the local density $\vr(x,y)$ is randomly modified. More
precisely, in one elementary time step, corresponding to the
``life-time'' of the shear band in a very loose granular sample, we
assume that the density along the shear band acquires random
uncorrelated values picked from a statistical distribution $p(\vr)$.
The uncorrelated character of the distribution is however justified
only on a mesoscopic scale.

After the elementary strain event, we have a new density map
$\vr(x,y)$. We now simply reiterate this procedure as long as
desired: Namely we identify the new path which minimizes $S({\cal
P})$, and update the value of the densities along this path randomly.
As the purpose of the present article is to illustrate some
statistical aspects of this dynamics, we do not try to mimic any
specific granular system by imposing a realistic density distribution
or initial correlations in $\varrho$. We will choose here a simple
uniform distribution between 0 and 1 for $p(\vr)$. The mean value and
variance of the distribution $p$ can be chosen arbitrarily, since a
translation and rescaling of $\vr$ does not affect the result.

A key assumption of our model which may appear as precluding the
occurrence of a slow evolution toward a critical state is the
selection of the density values within a shear band from uncorrelated,
smooth distributions. In fact, we will show below that, on the
contrary, a collective and purely statistical effect produces a slow
increase of the mean density over large strains. 

Our model is furthermore discretized on a regular square lattice. We
have looked at two different kinds of square lattices to check the
robustness of our results. In the first one the value of the density
$\vr$ is carried by the bonds. The orientation of the lattice is
chosen so that the principal directions lies at $\pi/4$ with respect
to the $(x,y)$ axis as shown in Fig. \ref{Fig_lattice} a). In the
other version, density values are assigned to the sites of a square
lattice. In this case the minimal path can be connected through the
next nearest neighbours too as shown on Fig. \ref{Fig_lattice} b).
Both square lattices give exactly the same results so in the following
we just refer to them as square lattice realizations. 

We mention here that a third type of lattice, the hierarchical diamond
lattice \cite{ber} was also studied. The numerical results are
surprisingly, essentially unchanged by the unusual topology of this
recursively constructed lattice. The easier construction of this
lattice however allows us to solve the model analytically thus giving
us a quantitative picture of the behaviour of the system. These
results are presented in \cite{TKKRhier} where the intimate
relationship of our model to other models of statistical physics is
also discussed.

\bfg
\centerline{\epsfig{file=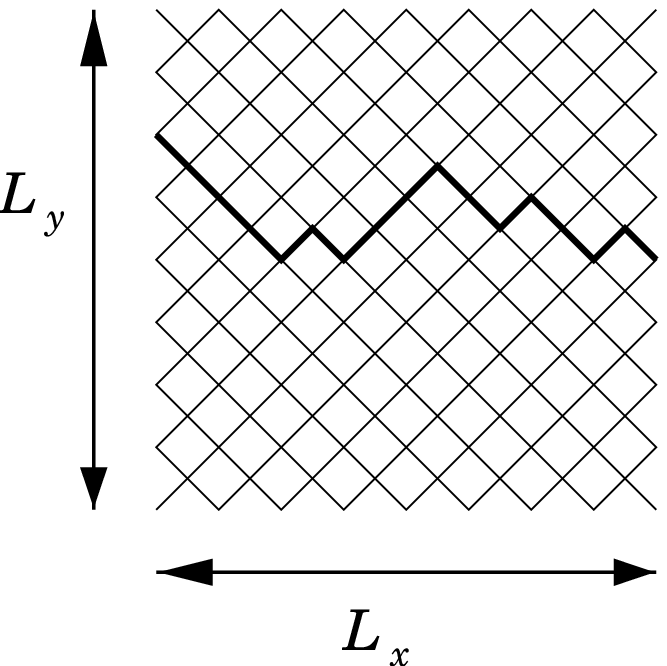,height=4truecm}
\epsfig{file=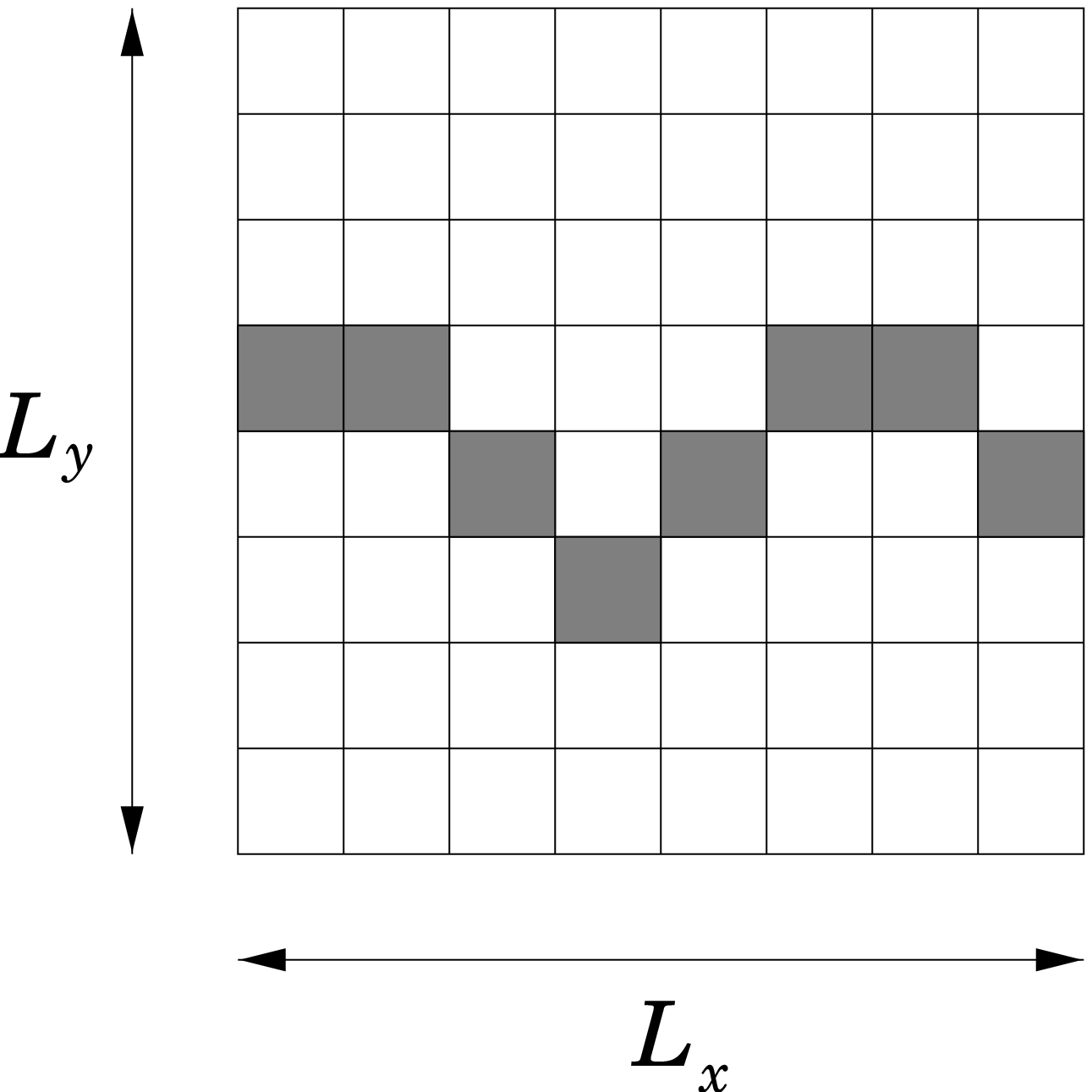,height=4truecm}}
\medskip
\caption{\label{Fig_lattice}\small
a) Visualization of the tilted lattice. The shear band is marked with a
thick line. b) A sample configuration of the minimal path on the
normal square lattice with densities assigned to sites.
}\efg

The rules of our model, finding the extremal directed spanning path at
every instant, is similar to finding the ground state of a directed
polymer in a random potential \cite{HalHeaRev}. However, in our case
this potential is uncorrelated only at the beginning; it changes in
time through the process described above, of ascribing new densities
to all sites along the minimal path. It is clear from this
relationship between the models, that the shear band is self-affine
with a Hurst exponent $\zeta=2/3$ at the beginning, i.e. the
transversal fluctuations of the band grow with the size ($L_x$) of the
sample width as $L_x^\zeta$. We will discuss the time evolution of the
roughness later in this paper.

%%%%%%%%%%%%%%%%%%%%%%%%%%%
\section{Numerical results}
%%%%%%%%%%%%%%%%%%%%%%%%%%%

\bfg
\noindent\centerline{\epsfig{file=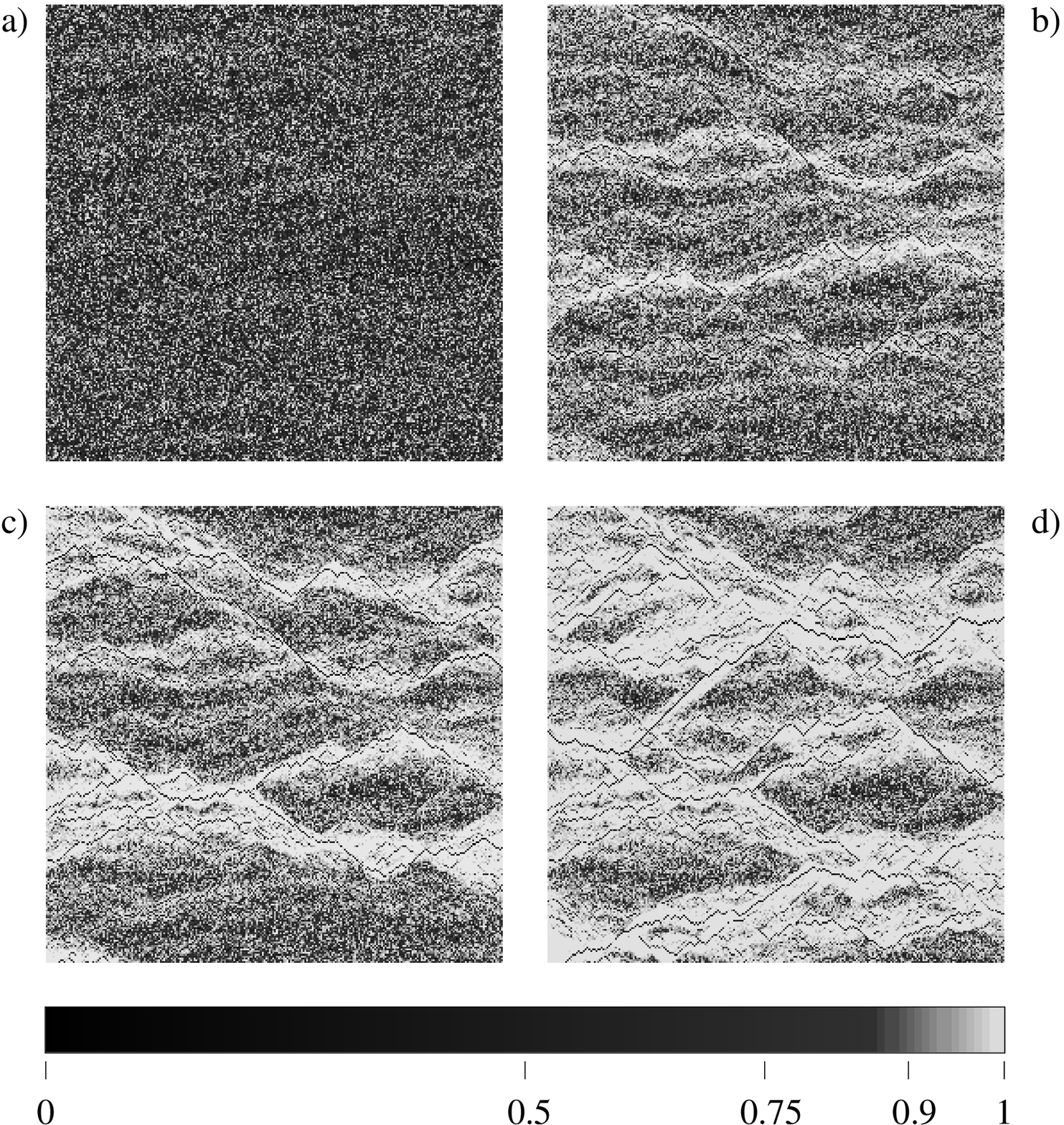,width=9truecm}}
\medskip
\caption{\label{Fig_maps}\small
Snapshots of densities of a square lattice of size $256$ by
$256$ at time a) $1000=4L$, b) $10^4=40L$, c) $10^5=400L$ and d)
$10^6=4000L$. The gray scale is presented at the bottom. The actual
shear band is drawn in black.
}
\efg

We first show the density map, Fig.~\ref{Fig_maps}, of the system at
different times, $t/L$ ranging from 4 to 4000. The grey scale chosen
focuses on the vicinity of 1 so as to highlight the progressive
densification. It may appear counter-intuitive at first that the rest
of the medium shows a densification at all, when the only dynamics
consists of finding a minimal path and updating sites along it
randomly. However the reason is simply that this update systematically
hunts out the sites with the lowest density values and replaces them. 

At early times, we observe an apparently uncorrelated field. However,
as time proceeds, it is possible to distinguish preferential channels
of high density aligned along the direction ($x$-axis) of the minimal
paths. These channels however have a significant width which shows
that though the minimal path has been confined to this zone, it has
enough freedom to explore different neighboring configurations and
achieve a significant local densification. We also see within these
wide and dense channels, a single path with a lower density. This has
been the last active minimal path in the channel. As time passes, the
number of channels increases, and so does their width. They get partly
interconnected, leaving always the same scars of low density paths.
Finally, at the latest time shown on the figure, the average density
is quite high, and traces of ancient minimal paths are still visible.
Nevertheless what is striking is the occurrence of islands entrapped
by these high density channels, where the density map looks like at
the very early stage of the time evolution. This signals that these
regions have basically not been visited by the minimal path during the
entire history of the system. These features are quite generic, and
they reveal that the spatial (and temporal) organization of the
activity is rather complex. The rest of the study is devoted to a more
quantitative account of this activity, of the resulting kinetics of
compaction, and the unexpected finite size effects which appear in
this problem. 

In the following subsection, we will introduce the main measurements
performed numerically on the model.

%%%%%%%%%%%%%%%%%%%%%%%%%%%%%%%%%%%%%%%%%%%%%%%%%%%%%%%%%%%%
\subsection{Definitions of numerically measured quantities}
%%%%%%%%%%%%%%%%%%%%%%%%%%%%%%%%%%%%%%%%%%%%%%%%%%%%%%%%%%%%
%%%%%%%%%%%%%%%%%%%%%%%%%%%%%%%%%%
\subsubsection{Average density} %
%%%%%%%%%%%%%%%%%%%%%%%%%%%%%%%%%%

The most important quantity is the {\it average density} of the sample
that we define as the mean of the density of the inactive sites, i.e.
the sites {\it not belonging} to the shear band. We denote this by
$\avrho$. This definition is convenient because $\avrho$ increases
monotonically by the rules of our model. In experiments one of the
most frequently measured quantities is the volumetric strain which is
just the change in the inverse of the average density. In our model,
when $p(\vr)$ is chosen to be the uniform distribution in the interval
$[0,1]$, the average density is bounded by ($\avrho\leq1$). It is easy
to see that in finite systems the steady state (the asymptotics)
cannot be other than a system with maximal densities everywhere except
for a path which will be always chosen as the minimal path. This
state is equivalent to $\avrho=1$.

We are interested in the approach to this asymptotics so we plot the
quantity $(1-\avrho)$ as a function of time. Within the granular
medium context, this means that we mainly study a loose initial state
and its convergence to the critical state \cite{Critstate1,Critstate2}.
However, we will also present results obtained when one starts from a
high initial density later (Section~\ref{sechidens}).

Figure \ref{Fig_avrho} shows that the difference of the average
density from its asymptotic value first remains almost constant during
a first stage $t/L\ll 1$, and then it decreases steadily. This first
increase of $\avrho$ is well captured by a reduced time equal to
$t/L$. However, as time progresses, the average density increases more
and more slowly. Quite strikingly, the larger the system size, the
slower the increase in density. Further down this will be interpreted
as a breakdown of ergodicity.

\bfg
\centerline{\epsfig{file=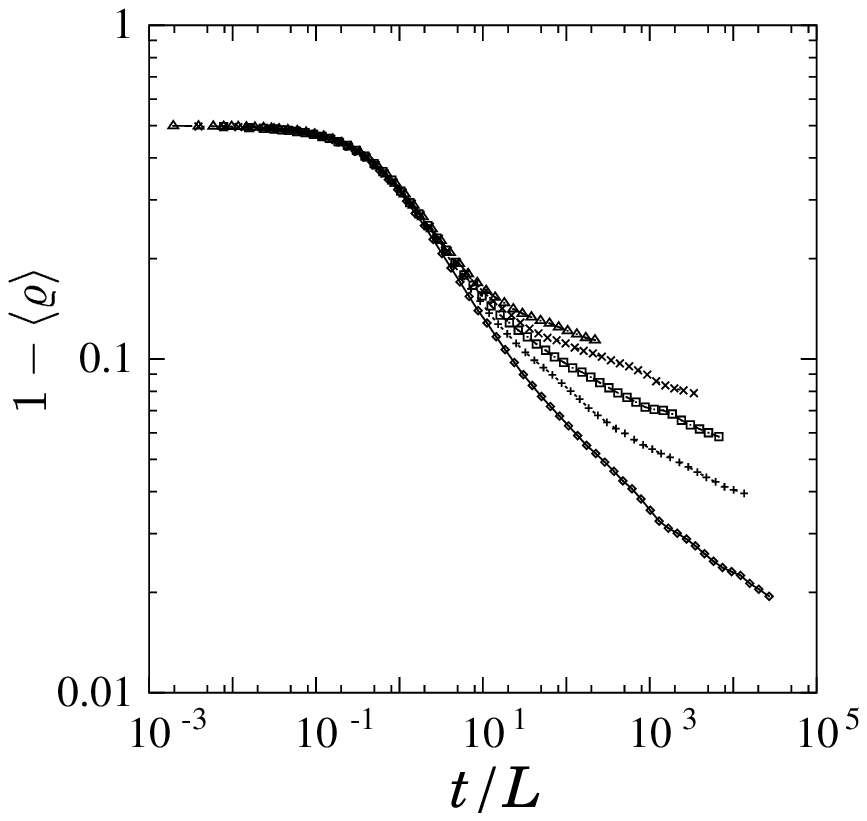,height=6truecm}}
\caption{\label{Fig_avrho}\small
The difference of the average density from its
asymptotic value $1$ is plotted as the function of time. The system
sizes are $L=32$, $64$, $128$, $256$, $512$. The average was done over
the inactive sites on the lattice and for an ensemble of $20$ to $1000$
samples.}
\efg

%%%%%%%%%%%%%%%%%%%%%%%%%%%%%%%%%%
\subsubsection{Shear band density}
%%%%%%%%%%%%%%%%%%%%%%%%%%%%%%%%%%

It is also natural to define the {\it density of the shear band} that
we denote by $\vrSB$. This is just the average density of the sites
along the minimal path (before updating them). As already mentioned in
the introduction, we assume that the maximal static shear force is a
single function of the density. Thus the density of the shear band can
be related to the shear stress in experiments.

Figure~\ref{Fig_rho_SB_t} shows the evolution in time of the
difference $(0.5-\vrSB)$. As expected, as time proceeds, the density
along the shear band will tend toward the average of the random
densities which are used to refresh the sites or bonds along the shear
band. Using a uniform distribution of densities between $0$ and $1$
implies that this average is $0.5$. Note that in contrast to the
previous case, the reduced time $t/L$ accounts nicely for the time
evolution of this quantity for all system sizes for $t/L \le 10^4$.
This is a second puzzle we will try to address further in the
following sections as well as in ~\cite{TKKRhier}.

\bfg
\centerline{\epsfig{file=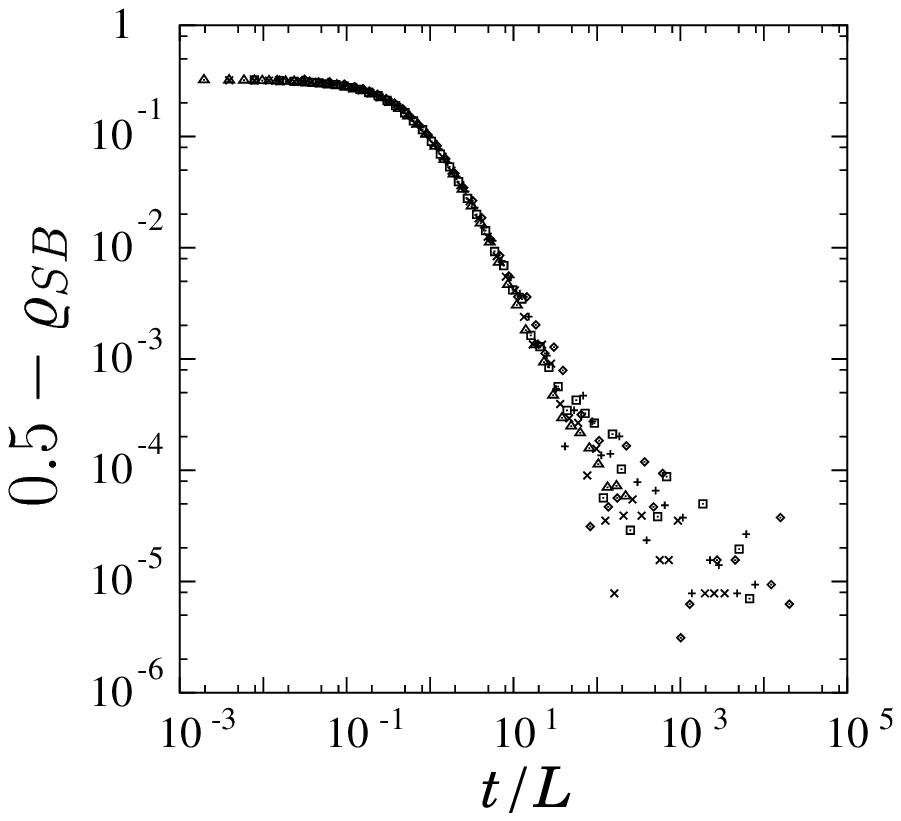,height=6truecm}}
\caption{\label{Fig_rho_SB_t}\small
The difference of the mean density of the shear band from its
asymptotic value $0.5$ vs. time. Notation and system sizes are the
same as in Fig. \ref{Fig_avrho}.
}
\efg

%%%%%%%%%%%%%%%%%%%%%%%%%%%%%%%%%%%%%
\subsubsection{Mean Hamming distance}
%%%%%%%%%%%%%%%%%%%%%%%%%%%%%%%%%%%%%

We also calculate the {\it Hamming distance} of two successive shear
bands which is defined as the number of sites (or bonds) by which two
consecutive shear bands differ. We denote this distance by $d$. The
natural normalization is to divide this distance by the total length
of the path, $L_x$. As we will see this quantity is very useful in
characterizing the time evolution of the localization process.

Fig.~\ref{Fig_diff_t} shows that the mean Hamming distance is close to
unity (i.e. two consecutive paths do not overlap at all) at early
times, and decrease towards $0$ for $t/L\gg 1$. We recall that when
the distance is equal to $0$, then the two consecutive conformations
of the shear band are identical, in spite of the total renewal of
random densities along them. This indicates that the shear bands have
a tendency to remain more and more persistent as the system ``ages''.
We will analyze further the complete statistical distribution of the
Hamming distance later in this paper.

\bfg
\centerline{\epsfig{file=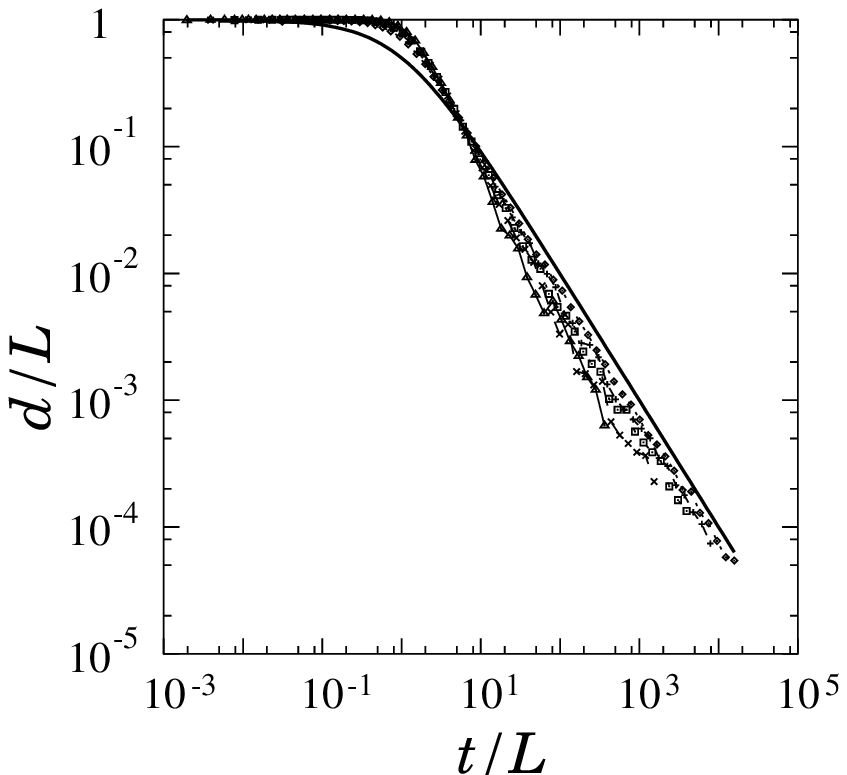,height=6truecm}}
\caption{\label{Fig_diff_t}\small
The average Hamming distance versus time. The same system sizes were
scaled together as on Fig. \ref{Fig_avrho}. The analytical prediction
\cite{TKKRhier} $1/(t+1)$ is plotted over the data. 
Note that scaling
with system size displays systematic corrections for larger systems.
}
\efg

%%%%%%%%%%%%%%%%%%%%%%%%%%%%%%%%
\subsubsection{Cumulative shear}
%%%%%%%%%%%%%%%%%%%%%%%%%%%%%%%%

\bfg
\centerline{\epsfig{file=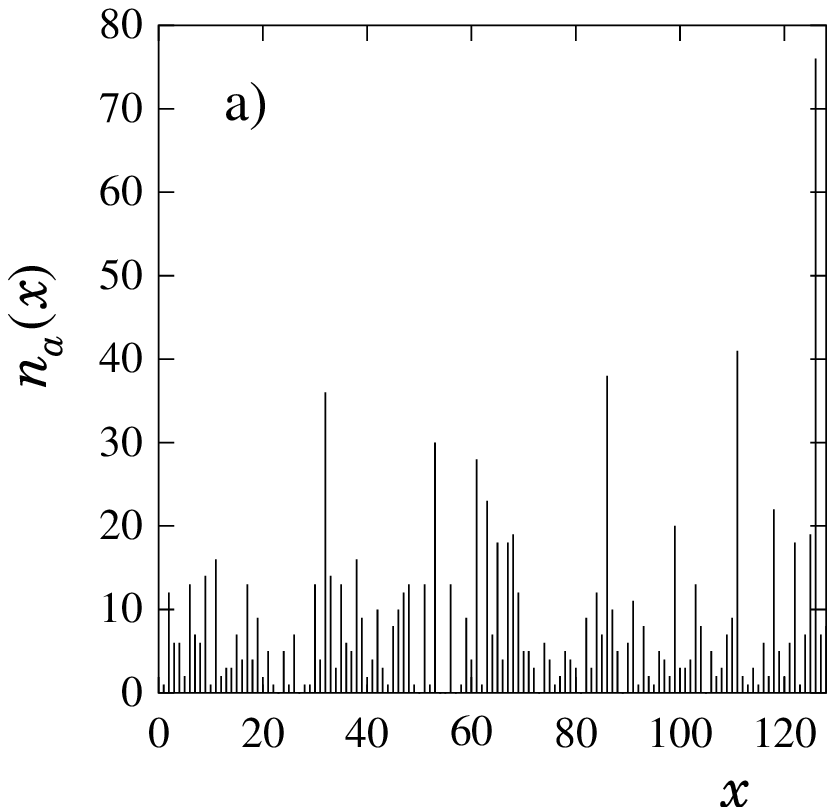,height=6truecm}}
\medskip
\centerline{\epsfig{file=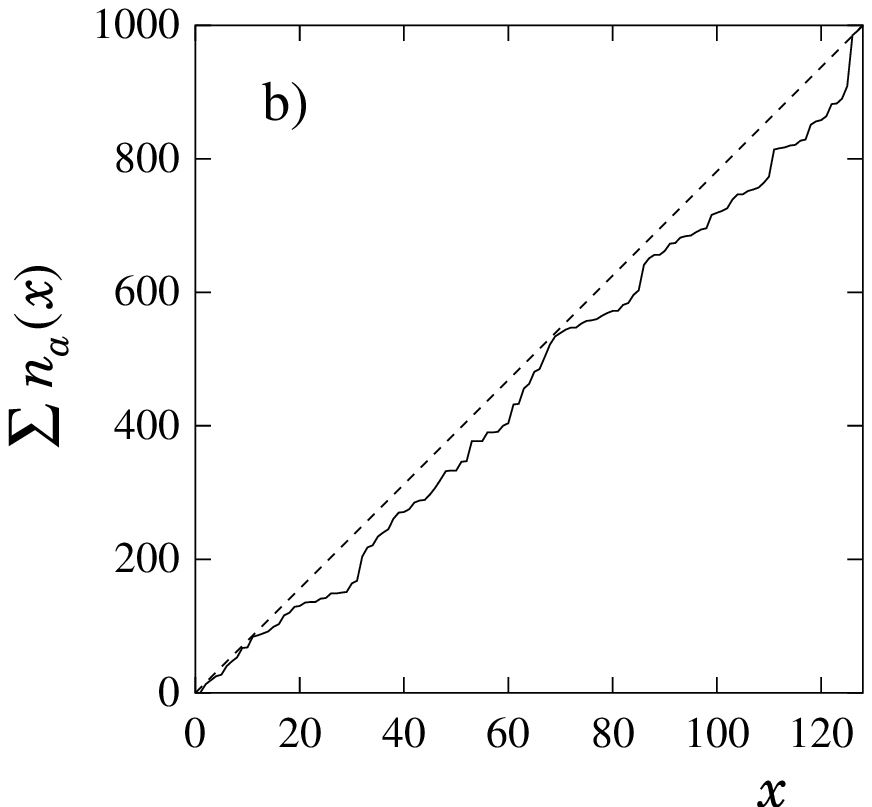,height=6truecm}}
\medskip
\caption{\label{Fig_sampleage}\small
Numerical Estimation of the cumulative shearing $\sigma_{cum}$. Figure
a) represents the number of times ($n_a$) a site $y$ was active up to
time $t=1000$ in a cross-section of a $128$ by $128$ sample. Figure b)
is the cumulative representation of a): $\sum_{j=0}^y n_a(j)$. The
dashed line indicates the homogeneous case.
}
\efg

An experimentally relevant quantity is the {\it cumulative shearing}
denoted by $\sigma_{cum}$. The numerical procedure we follow to obtain
this quantity in our model is the following (see Fig.
\ref{Fig_sampleage}): We mark a line in the $y$-direction (see Fig.
\ref{Fig_lattice}). We measure the total activity $n_a(y)$ along the
line, {\it i.e.} the number of instances when the shear band went
through a point $y$ on the line. From this, we define
\be
\sigma_{cum}(y)=\sum_{j=0}^y n_a(j).
\ee
By definition $\sigma_{cum}(L_y)=t$, since at every instant, the shear
band has necessarily to pass through one or the other site on a cut
along the $y$ axis. The fluctuations of $\sigma_{cum}(y)$ about its
mean value $(t/L_y)y$ then reflect
the inhomogeneity of the shear process within the sample. In Fig.
\ref{Fig_cumshear}, we track the time evolution of $\sigma_{cum}(y)$,
after subtracting out the mean value. As indicated, a snapshot of this
quantity encodes the history of the process of shearing in this
system.

\bfg
\centerline{\epsfig{file=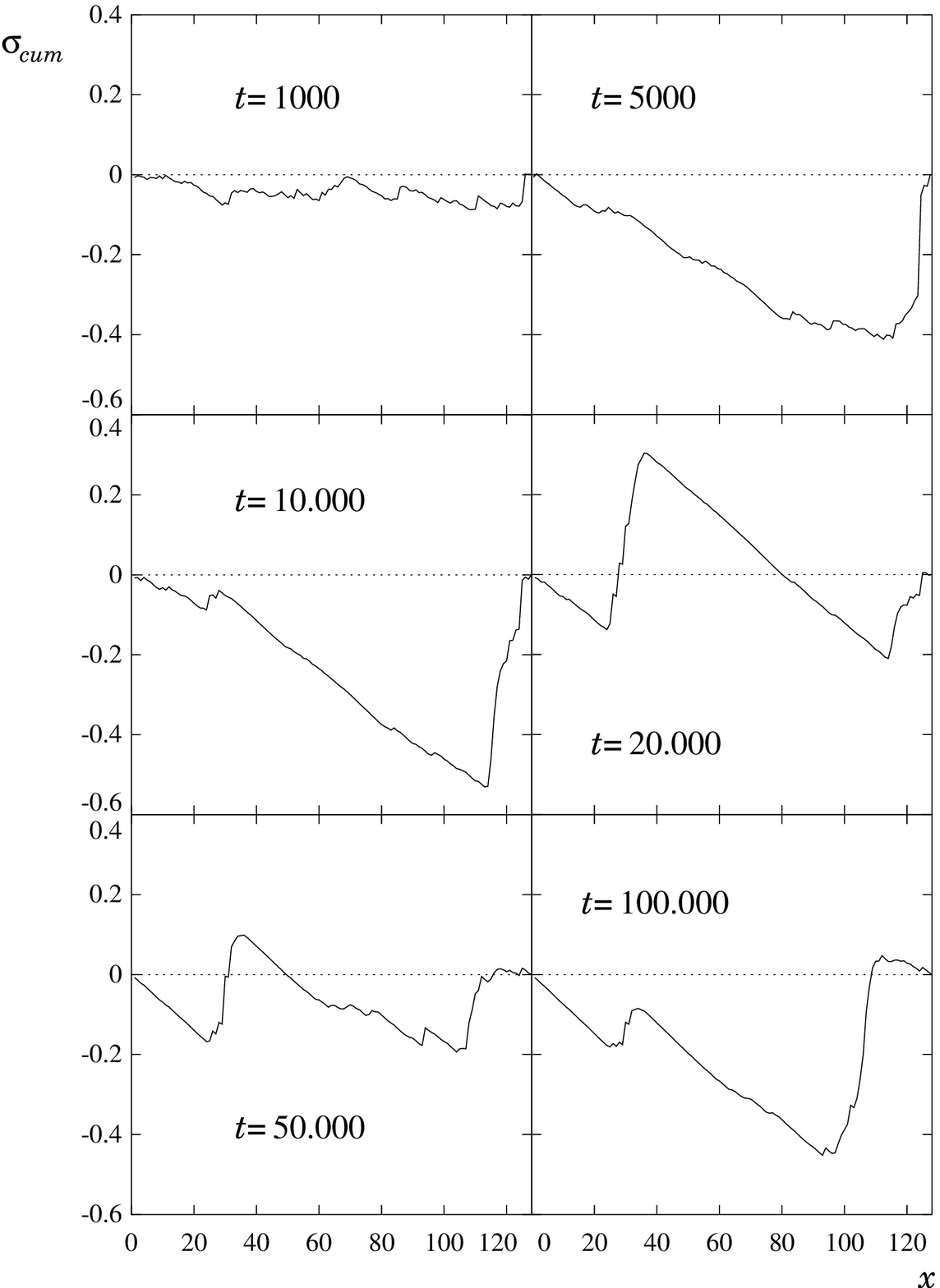,width=8truecm}}
\medskip
\caption{\label{Fig_cumshear}\small The cumulative shear corrected by
the average displacement for a system of size 128 by 128. As one can
observe, this quantity encodes the history of the process; a dip in
the profile indicates the presence of the shear band and the depth of
the dip is indicative of the amount of time it has spent in that
location. For example, in the beginning when successive shear bands
are distinct, every site is visited approximately equally and the
profile has no deep peaks or valleys. After this, at $t=5000$, the
first shear band gets localized at around $x=100$. This lasts until
about $t=20,000$, and then it jumps to $x=40$. After spending some
time here it jumps back, close to its previous position
($2~10^4<t<5~10^4$). } 
\efg

%%%%%%%%%%%%%%%%%%%%%%%%%%%%%%%%
\subsection{Early time regime}
%%%%%%%%%%%%%%%%%%%%%%%%%%%%%%%%

It is apparent from Figures \ref{Fig_avrho}, \ref{Fig_rho_SB_t} and
\ref{Fig_diff_t} that the initial behavior of the model is very
different from the late stages. In the former regime, the average
distance $d$ (Fig. \ref{Fig_diff_t}) is equal to the system size
indicating that successive shear bands do not overlap at all. In other
words there is an effective strong repulsive interaction between them.

The density on the shear band (Fig. \ref{Fig_rho_SB_t}) provides an
explanation for this behavior. From the directed polymer
\cite{HalHeaRev} picture it is known that the first shear band has a
mean density of $\vrSB(t=0)=e^*\approx 0.22$ on the tilted square
lattice. The state of the shear band after the densities of the sites
along it have been changed, is independent of its previous state and
the mean density is now $0.5$ (since it is just an average of $L_x$
independent random numbers taken from the uniform distribution [0,1]).
This implies that at the next instant, the chosen shear band will have
a very low probability of sharing bonds/sites with the previous one,
since there will still be many spanning paths with density smaller
than $0.5$. Thus, until all the sites are visited at least once,
successive shear bands have few sites/bonds in common ($d=L_x$ in
Figs. \ref{Fig_diff_t}). Also in this regime, since the path sweeps
all sites through the cumulative shearing appears to be homogeneous
(Fig. \ref{Fig_cumshear} first plot). As mentioned in the
introduction, this uniform shear strain is consistent with the
experimental observations that no well-defined shear-bands persist
over observable time scales in loose samples.

The characteristic time to build up correlations is set by the
sweeping through the sample, i.e., it is given by $L_y$. This is the
reason why in the early time regime the plots in Figs. 4-6 can be
scaled together with $L_y$.
 
%%%%%%%%%%%%%%%%%%%%%%%%%%%
\subsection{Localization}
\label{Sec_local}
%%%%%%%%%%%%%%%%%%%%%%%%%%%

The above described behavior is drastically changed as time goes on
and the shear band gets localized for very long times at the same
location. 

As the average density increases, the probability of choosing a
minimal path with density less than $0.5$ starts decreasing, and thus
for the new path, it becomes more favorable to overlap to a greater
and greater extent with the previous one. As a result, the density of
the region where the path is located, is further increased, hence
trapping the minimal path in canyon-like structures surrounded by
extremely high density regions (see Fig. \ref{Fig_maps} b-d). 

The early repulsive interaction is thus now inverted to an attractive
one as is shown by the rapid $1/t$ like decrease of the Hamming
distance (Fig. \ref{Fig_diff_t}), as well as the decrease of $0.5 -
\vr_{SB}$ and of $ \avrho$ from its initial value (Figs.
\ref{Fig_rho_SB_t} and \ref{Fig_avrho} respectively) in this regime.

There are a number of consequences of the localization. First, as time
goes on, the shear band (which was not visible at all in the density
map of the system (see Fig. \ref{Fig_maps} a)) becomes more and more
apparent until finally it becomes localized at a given position for
macroscopic times. This is in accordance with experiments and with the
critical state concept \cite{Critstate1,Critstate2}. Secondly, since
the same path for the shear band is chosen most of the time, its
density saturates to its asymptotic value $0.5$ (Fig.
\ref{Fig_rho_SB_t}). Again this is consistent with the experimental
observation that the density within the shear band tends to achieve a
well defined value somewhat smaller that the rest of the medium for
dense granular media. Simultaneously, the shear stress saturates to a
constant value since this is imposed by the shear band itself
\cite{Wood}. However, we note that in our model the global density of
the system continues to increase in time, albeit very slowly. This
extremely slow trend may well be out of reach experimentally.
However, it is known \cite{Wood} that the shear stress saturates much
faster than the volumetric strain (the average density in our case)
which is clearly justified by the numerical results.

On the cumulative shear which is a straight line with small
statistical fluctuations in the early time regime, there appear
step-like structures, indicators of progressively more persistent
localization (Fig. \ref{Fig_cumshear}). However, this localization is
not everlasting since the shear band may perform big jumps to other
local minima. This can be seen on the series of cumulative shear
curves in the form of certain steps disappearing and others becoming
more prominent. This prediction could easily be tested experimentally.

%%%%%%%%%%%%%%%%%%%%%%%%%%%%%%%%%%%%%%%%%%%%%%%%%%
\subsection{Systems with different aspect ratios}
%%%%%%%%%%%%%%%%%%%%%%%%%%%%%%%%%%%%%%%%%%%%%%%%%%

All changes take place along the shear band, which is aligned along
the $x$ direction, and thus we may anticipate that the $x$ and $y$
direction will play different roles. Therefore in this section, we
study the influence of the width and length of the system. In what
follows, we use the terms {\it long} for samples with $L_y<L_x$ and
{\it wide} in the opposite case ($L_y>L_x$). 

With very long and wide systems we are able to separate the two kinds
of dynamics described in Section \ref{Sec_local}. If one considers a
long sample ($L_y/L_x$ is small), we expect the time evolution to be
independent of $L_x$ for all quantities of interest since the lattice
can be split into subparts placed in series. Thus one may expect the
large jumps to disappear and the average density to scale solely with
$L_y$ (Fig. \ref{Fig_longdens}). A wide system, on the other hand,
might be expected {\it a priori} to behave like a number of
competing subsystems.

%%%%%%%%%%%%%%%%%%%%%%%%%%%%%%
\subsubsection{Long systems}
%%%%%%%%%%%%%%%%%%%%%%%%%%%%%%

\bfg
\centerline{\epsfig{file=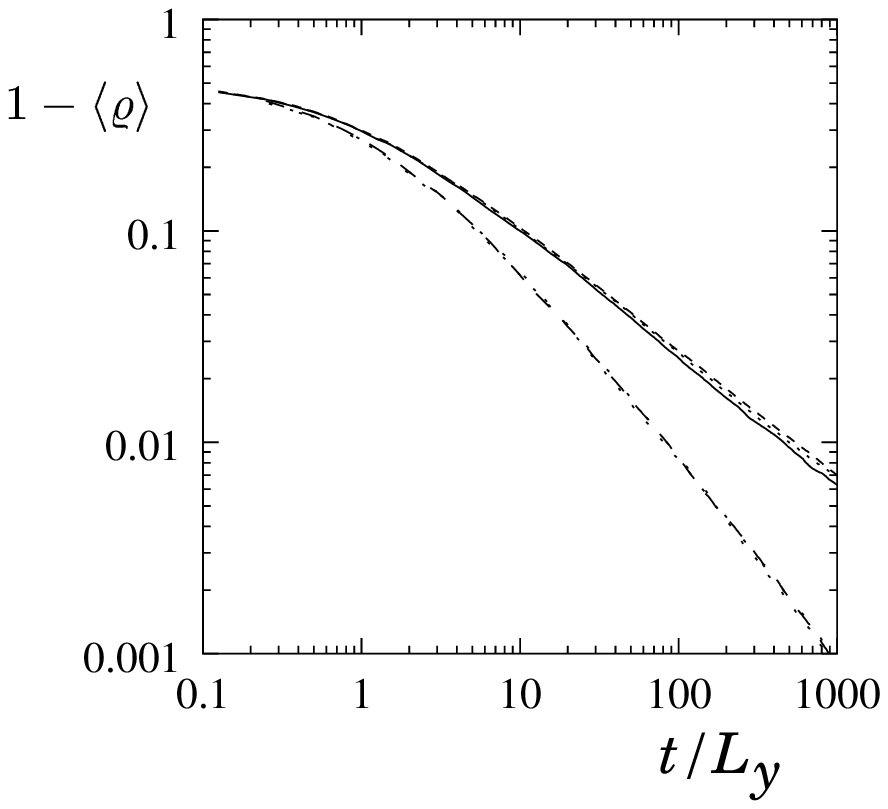,height=6truecm}}
\medskip
\caption{\label{Fig_longdens}\small
Long samples with width $L_y=4$ (lower curves) and $L_y=8$ (upper
curves). Three different lengths were used in both cases $L_x/L_y=2.5$,
$5$ and $10$. 
}\efg

On Fig. \ref{Fig_longdens} we have plotted the time dependence of the
density in long samples with $L_y=4$ (lower curves) and $L_y=8$ (upper
curves). The $t/L_y$ scaling is excellent in both cases. However, the
densification obeys a different time evolution for different $L_y$.
The rate at which the density increases is slower as the width
increases.

%%%%%%%%%%%%%%%%%%%%%%%%%%%%%%
\subsubsection{Wide systems: Breakdown of ergodicity}
%%%%%%%%%%%%%%%%%%%%%%%%%%%%%% 

Wide samples can be considered as subsystems placed next to each other
and coupled in parallel. In contrast with the previous case, we will
see that the evolution of the different subsystems cannot be accounted
for by a simple average. 

In the case of the wide systems the same plot as before (Fig.
\ref{Fig_widedens} a) shows no $t/L_y$ scaling for small system sizes.

\bfg
\centerline{\epsfig{file=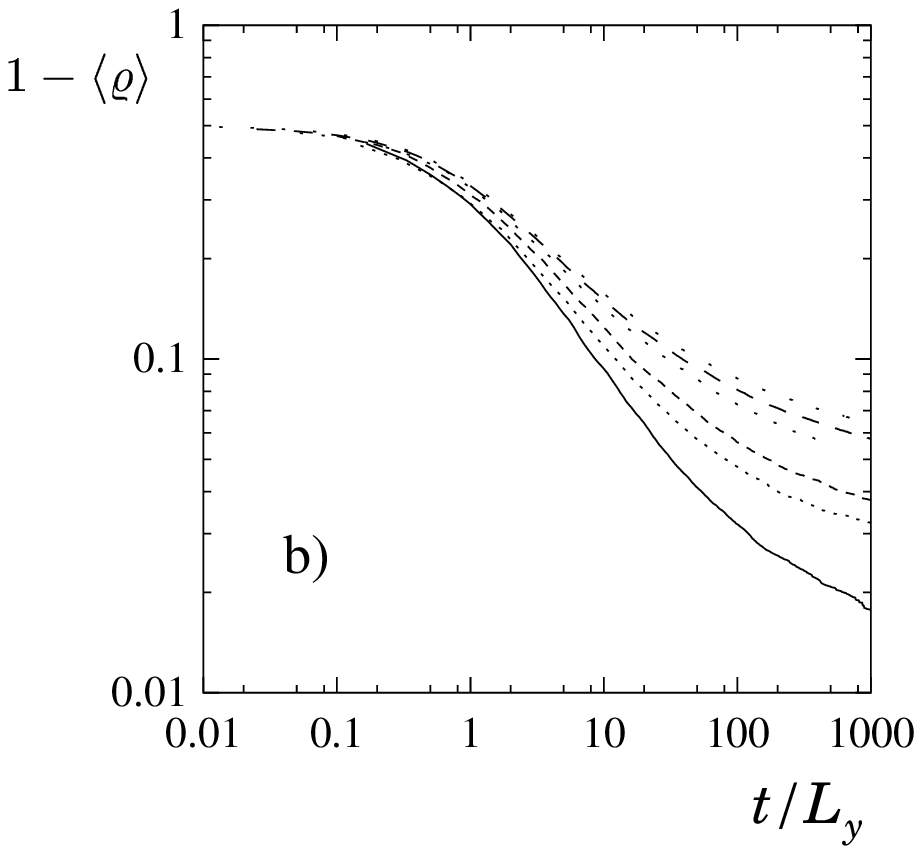,height=6truecm}}
\medskip
\caption{\label{Fig_widedens}\small Time dependence of the density
difference from its asymptotic value for wide samples. The length of
the system along the shear band is $L_x=5$ the widths are $L_y=5$, 10,
20, 40, 80 and 160 from bottom to top, respectively.
}\efg

We could imagine the following construction: suppose we split a given
wide system into two subsystems of size $L_x\times(L_y/2)$. Provided
$L_y$ is large enough, we can ignore the interaction between the two
subsystems, and thus we could study independently the time evolution
of both subsystems. Now, if we merge them again, we realize that the
only reason why the resulting densification could differ from the
measurement on the separate subsystems is that the time $t_1$ the
shear band has stayed in subpart $1$ is far from being equal to $t/2$.
In other words, the breakdown of the data collapse of $1-\avrho$ {\it
vs.} $t/L_y$ is a {\it breakdown of ergodicity}. This is naturally
associated to what we termed ``localization'' earlier.

In order to make this concept more explicit, let us consider an extreme
version of such a breakdown of ergodicity. During a first stage, up to
$t/L_y$ of order 1, the activity is evenly spread over the system. Then
we assume that after such a time, the activity remains confined in a
subsystem of size $L_x\times\ell$. All other subsystems are assumed
not to be visited by the shear band, and thus their density is
quenched at their value reached at the onset of localization,
$\varrho_0$, at time $t_0=\theta L_y$. The global density will thus
obey
\be
1-\avrho(t)={(1-\varrho_0)(L_y-\ell)+f_\ell(t-\theta(L_y-\ell))\ell
\over L_y}
\ee 
where $f_\ell(t)=1-\avrho(t)$ describes the densification of the
representative cell of size $L_x\times\ell$ where the activity is
confined, and thus $1-\varrho_0=f_\ell(\theta\ell)$.

In this crude scenario, we note that the global density does not
converge to $1$ as time goes to infinity, but rather remains stuck at
a value such that $(1-\avrho(t))\to f_\ell(\theta\ell)(1-\ell/L_y)$.
In more quantitative terms, we tried to carry out such a procedure,
and indeed for a fixed $L_x$, it is possible to account for the time
evolution of systems of different width using $\varrho_0$ as free
parameter and calculating $\ell$ from the $L_y$ dependence. It turns
out that $\ell$ changes for small values of $L_x$ but becomes constant
($\ell\simeq 30$) above the system size of $L_x\simeq 30$. The test of
this analysis can be seen on Fig. \ref{Fig_widescale}. However, the
asymptotic density turns out to depend on $L_x$.

The conclusion is that although, such an extreme modeling of the
localization is able to capture part of the strong size effects
observed numerically, it is too crude to provide a quantitative
account of the densification. The hierarchical lattice provides us with
a convenient case where an analytical investigation of this breakdown of
ergodicity can be made. It is shown in Ref. \cite{TKKRhier}, that the
local ``age'' distribution assumes a multifractal distrbution whose
spectrum can be computed exactly. This property can then be used to
provide an expression of the density evolution in time. 

\bfg
\centerline{\epsfig{file=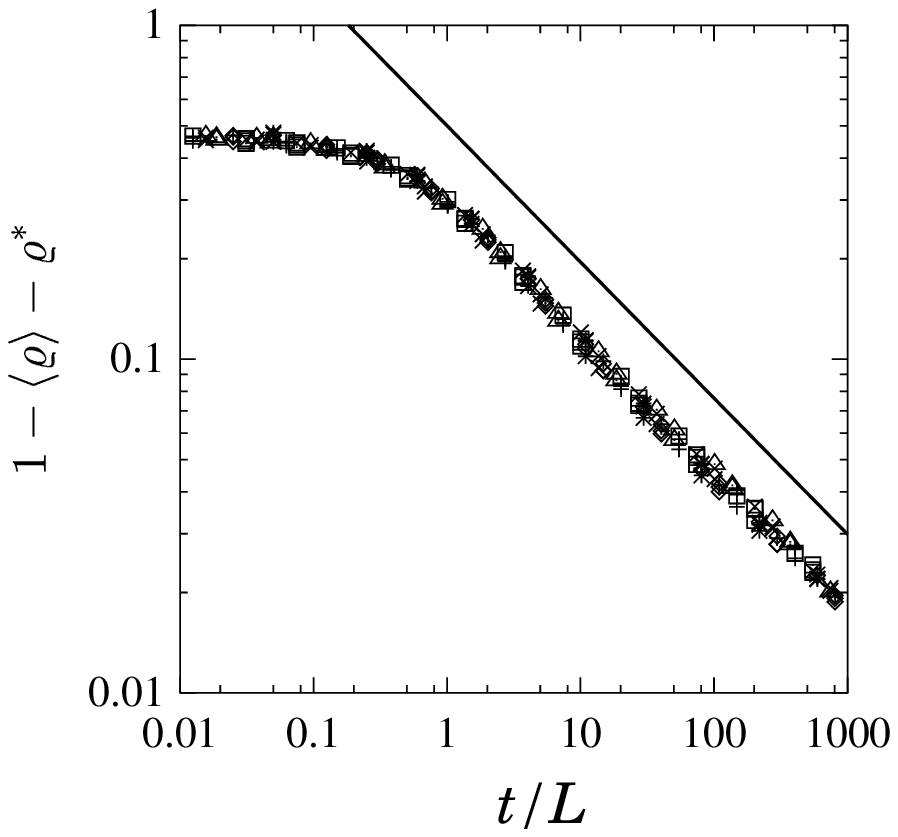,height=6truecm}}
\medskip
\caption{\label{Fig_widescale}\small
The same data as in Fig. \ref{Fig_widedens} with a different,
effective density $\vr^*=\vr_0(L_y-\ell)$. Note that the plot displays
nice scaling as well as a clean power law decay over at least three decades.
However if the simulation is continued further the average density
increases over $\vr^*$.
}\efg

%%%%%%%%%%%%%%%%%%%%%%%%%%%%%%%%%%%%%%%%%%%%%%%%%%%%%%%%%%%%%%%%
\subsection{Time evolution of the Hamming distance distribution}
%%%%%%%%%%%%%%%%%%%%%%%%%%%%%%%%%%%%%%%%%%%%%%%%%%%%%%%%%%%%%%%%

We study here the distribution of Hamming distances as a function of
time, $P(d,t)$ for both lattices. This quantity is the analogue of an
``avalanche distribution'', such as is usually studied in
self-organized critical(SOC) systems. As we will see below, this quantity
does indeed decay as a power-law like in {\it SOC} systems. However,
since a steady state is never reached, the power-law decay is
multiplied by a time-dependent prefactor. 

The distribution $P(d,t)$ is shown in Fig. \ref{Fig_disthist}. At
early times, this quantity is peaked around the maximum value
($d=L_x$) while in the localized regime, it becomes peaked at the
minimum value ($d=0$). This corresponds to the transition from
repulsive to attractive effective interaction between consecutive path
conformations.

\bfg
\centerline{\epsfig{file=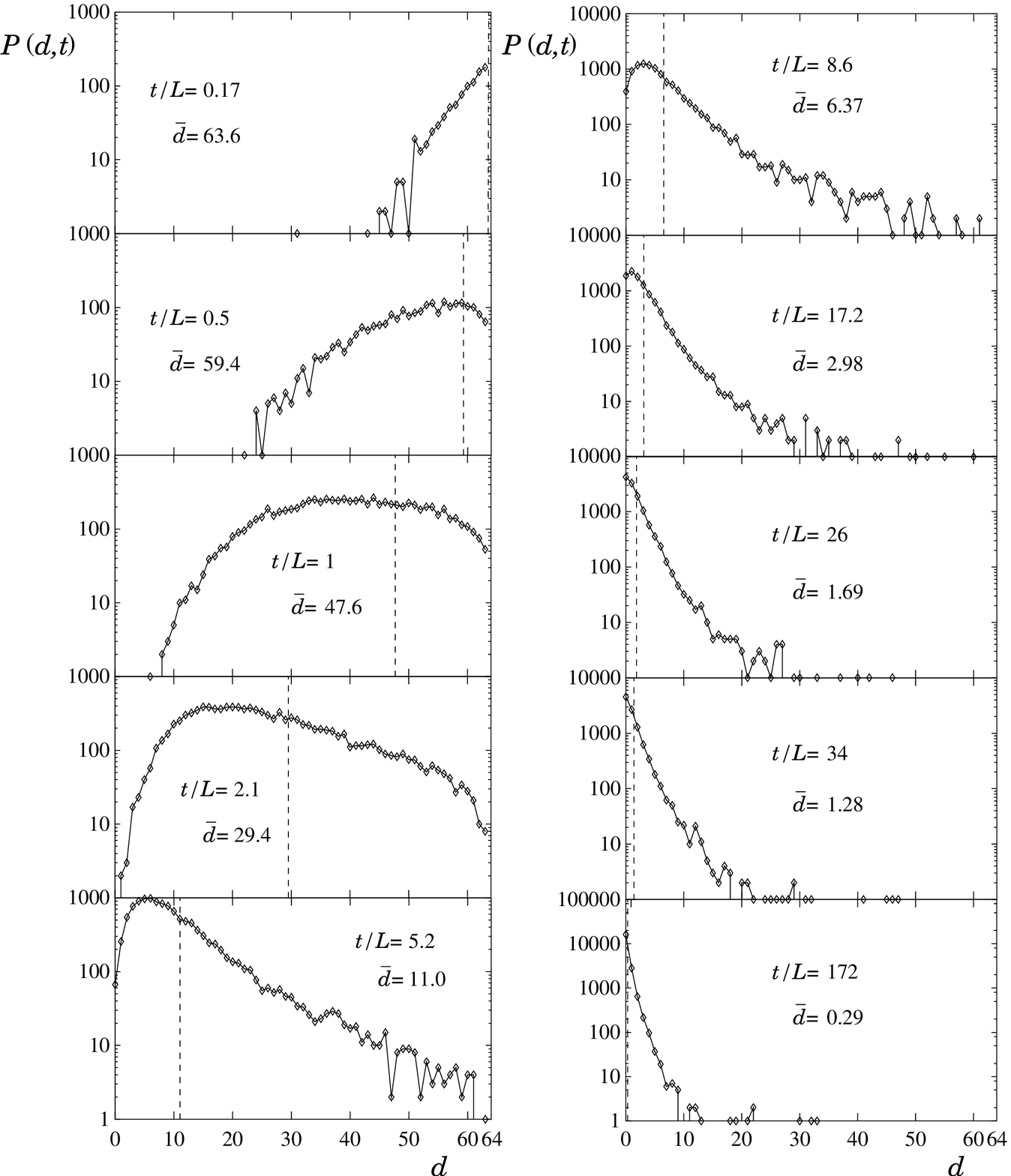,height=10truecm}}
\medskip
\caption{\label{Fig_disthist}\small
The distribution of the Hamming distance on a 64$\times$64 square system
for different 
times. The dashed lines indicate the mean ($\bar d$). 
}
\efg

At fixed (large) times, the distribution $P(d,t)$ decays as a
power-law of $d$, as can be seen in Figure~\ref{Fig_pdt}. The
measured exponent is 
\be
P(d,t)\propto d^{-3}
\ee
in addition to which there exists a peak at $d=0$ the amplitude of
which varies significantly with time. The decay of average $d$ as
$1/t$ (see Fig. \ref{Fig_diff_t}) found earlier, implies that the time
dependence of the $d\ne 0$ part is $p(d,t)\propto 1/(td^3)$. Thus,
including the different scalings with $L_x$ and $L_y$, we obtain
finally the asymptotic form
\be
P(d,t) \propto \frac{L_xL_y}{t d^{3}}
\ee
 
\bfg
\centerline{\epsfig{file=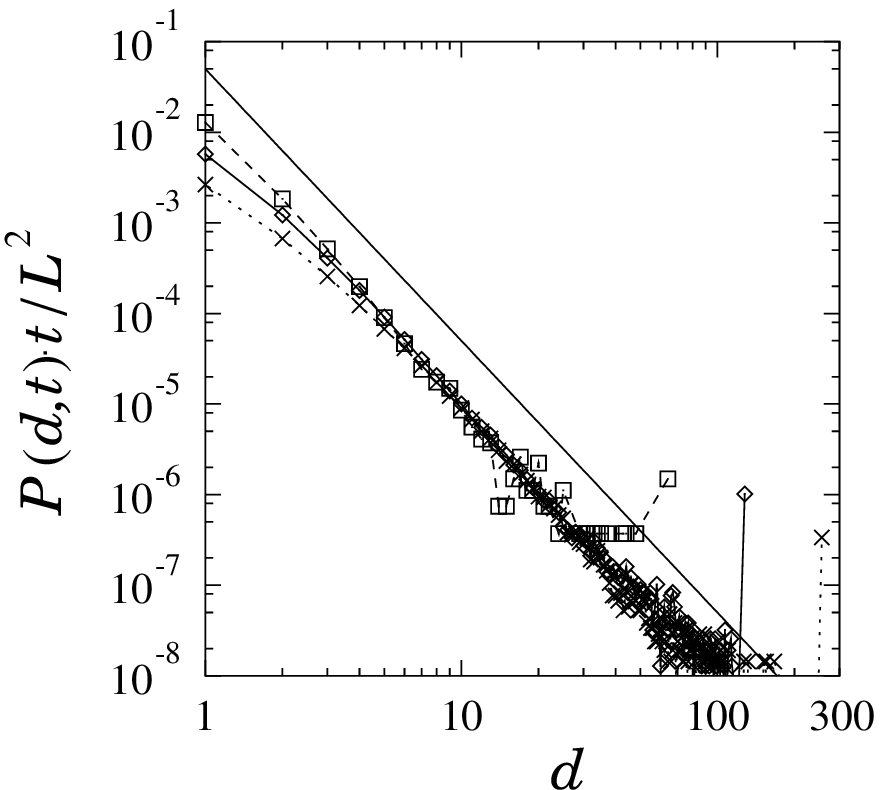,height=6truecm}}
\medskip
\caption{\label{Fig_pdt}\small
Scaling plot for the distribution of the Hamming distance $P(d,t)$ vs.
$d$. The three curves are for three square system of size: $L=64$ at
$t=10000$, $L=128$ at $t=1000$ and $L=256$ at $t=500$. The straight
line has slope $-3$ indicating that the decay of $P(d,t)$ with $d$ is
a power-law. Jumps of order of half the system-size however, seem to
have an enhanced probability.
}
\efg
Here again, the hierarchical lattice allows us to compute this
distribution analytically (for $L_x=L_y$), and we find that a similar
behavior is obtained \cite{TKKRhier}.

%%%%%%%%%%%%%%%%%%%%%%%%%%%%%%%%%%%%%%%%%%%%%%%%%
\subsection{Roughness exponent of the shear band}
%%%%%%%%%%%%%%%%%%%%%%%%%%%%%%%%%%%%%%%%%%%%%%%%%

From the directed polymer analogy we know that the shape of the shear
band is self-affine with an exponent of $\zeta=2/3$ for infinitely
large systems. It is an interesting question whether this property of
self-affinity is conserved in the time evolution of our system. We
have investigated this question and have found self-affine scaling for
all times, albeit with a time dependent Hurst exponent (Fig.
\ref{Fig_supr_aff}).

\bfg
\centerline{\epsfig{file=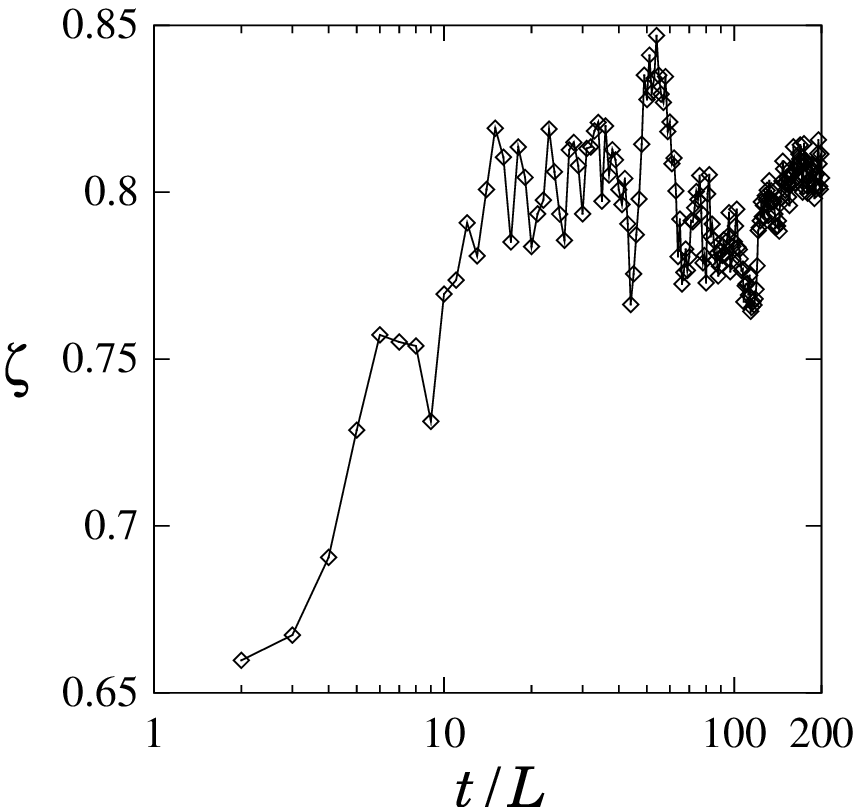,height=6truecm}}
\medskip
\caption{\label{Fig_supr_aff}\small
Estimated self-affine exponent as a function of time obtained
through Eq. \ref{Eq_zetainf} for system sizes varying from $L=64$ to
$L= 512$ .
}
\efg

We have estimated the value of $\zeta$ by measuring the width of the
shear band, $w_L(t)$, for different system sizes, $L\times L$. The
width is defined as the standard deviation of the $y$ coordinate of
the active path. The latter is expected to scale as $w_L(t)\propto
L^\zeta$ for a self-affine object. To estimate $\zeta$, the roughness
exponent, we have computed the ratio of two such widths for lattice
sizes differing by a factor of 2, and used the following estimate
\be\label{Eq_zetainf}
\zeta= {\log\left({w_{2L}(t)/ w_{L}(t)}\right)\over \log(2)}
\ee
where $w_{L}(t)$ is the width of a shear band in a $L\times L$ lattice
at time $t$. The results obtained for the tilted square lattice can be
seen on Fig. \ref{Fig_supr_aff}. It starts from $\zeta(t=0)=2/3$ as
expected from the directed polymer result and has an asymptotic value
of $\zeta\simeq 0.8$.

%%%%%%%%%%%%%%%%%%%%%%%%%%%%%%%%%%%%%%%%%%%%%%%%%%%%%%%%%%%
\subsection{Systems with high initial densities; relevance of initial
conditions}
\label{sechidens} 
%%%%%%%%%%%%%%%%%%%%%%%%%%%%%%%%%%%%%%%%%%%%%%%%%%%%%%%%%%%

\bfg
\centerline{\epsfig{file=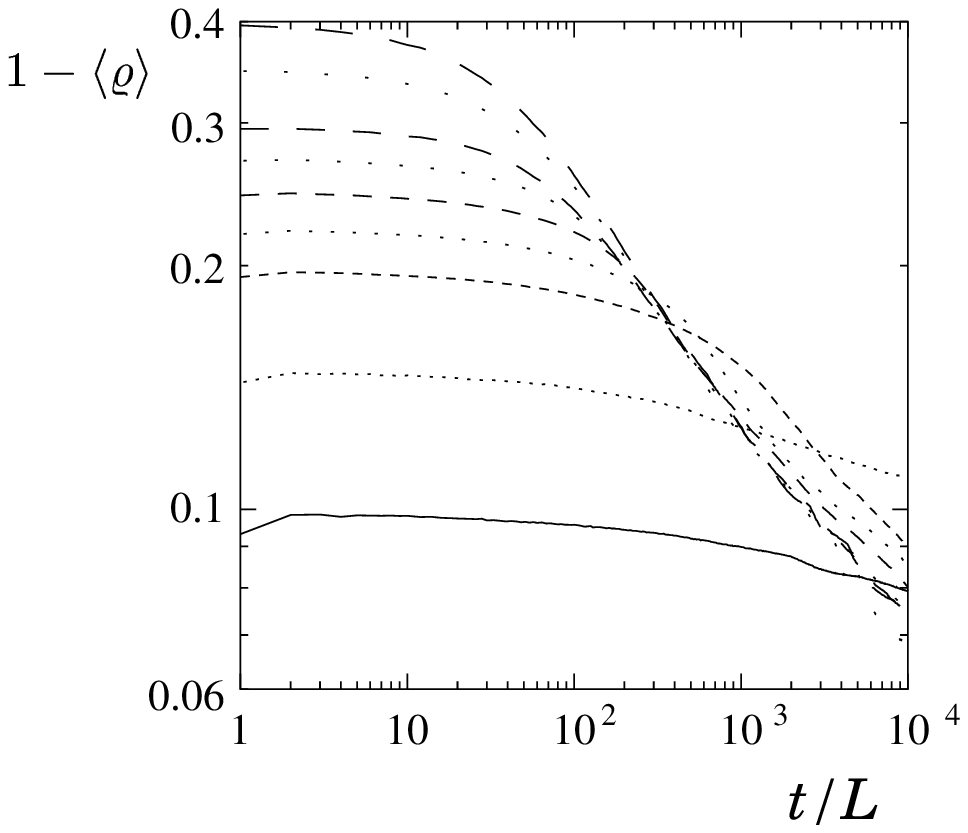,height=6truecm}}
\medskip
\caption{\label{Fig_high}\small
Time dependence of the difference of the average density from its
asymptotic value for starting densities with initial density ranges
$[\vr_{init}:1]$ from top to bottom respectively: $\vr_{init}=0.2$,
$0.3$, $0.4$, $0.45$, $0.5$, $0.55$, $0.6$, $0.7$, $0.8$. The system
size is $64$.
}\efg

We have so far studied the situation when a very loose granular medium
compactifies under shear, while simultaneously a shear band gets
quasi-localized in the system. It is also of interest to study
samples with higher initial density where it is known experimentally
that a shear band is localized from the very beginning. In this
section we study the interesting crossover to that state from the
previously described dynamics.

We choose initial densities from the interval $[\vr_{init}:1]$ with a
uniform distribution and varying $\vr_{init}$. In Fig. \ref{Fig_high}
the time evolution of the difference of the average density from its
asymptotic value is plotted using different initial conditions. Since
all previous arguments hold we assume that in finite systems the
asymptotic value of the average density is $1$ irrespective of the
initial density.

The striking result of these simulations is that if $\vr_{init}<0.5$,
all curves coincide in the decreasing regime. However, if
$\vr_{init}>0.5$ a different time evolution is observed for large
times. Thus we can assume that in the early time regime, when the
shear band is swapping uncorrelatedly, it visits all sites that have a
value less than the expectation value of the refreshing density
distribution $= 0.5$. After the first regime, as all small values are
eliminated, the system effaces the initial condition almost entirely.

On the other hand, starting from initial conditions with
$\vr_{init}>0.5$ we largely eliminate the possibility of big jumps.
The shear band fluctuations are now very small, involving changes in a
very few sites, and hence, the density change of the sample
is extremely slow. 

%%%%%%%%%%%%%%%%%%%%%%%%%%%%%%%%%%%%%%%%%%%%%%%%%%%%%%%%
\subsection{Summary of the numerical results}
%%%%%%%%%%%%%%%%%%%%%%%%%%%%%%%%%%%%%%%%%%%%%%%%%%%%%%%%

We have seen that the system densifies with time so as to approach a
unit density, i.e. the maximum available density from the distribution
used to refresh the sites. The kinetics of the densification is slow
(slower than any power-law). Moreover, after a first transient where
the reduced time $t/L$ accounts for the $L$ dependence, the compaction
process depends on the system size in a non-trivial way. The width of
the system is the parameter which really controls this anomalous
behavior, signaling that the competition between parallel paths may
somehow play a key role in this breakdown of system size rescaling.
This competition is a subtle one however, lying somewhere in between
complete localization of the path (which, as we saw in section IV B,
is too crude to mirror the actual scenario) and complete
delocalization (which, as mentioned earlier, accounts only for the
early time behaviour). 

The density maps display an interesting organization of
``canyon-like'' paths with density much lower than their immediate
surroundings (where the density approaches 1 quite uniformly).
Moreover, large regions are left quiescent, being systematically
avoided by the minimal paths. This contrast of high and low activity
within the same system is at the heart of the breakdown of ergodicity
observed after an initial transient. As remarked however, the
distribution of Hamming distances between consecutive paths displays a
somewhat simpler behavior where the role of the width and the length
of the system can be simply accounted for. 

%%%%%%%%%%%%%%%%%%%%%%%%%%%%%%%%%%%%
\section{Conclusion and discussion}
%%%%%%%%%%%%%%%%%%%%%%%%%%%%%%%%%%%%

Though very simply defined, our model seems to capture some essential
features of granular shear and provides at the same time several
predictions. The model demonstrates the self-organized mechanism of
the localization of the shear band in loose granular materials. As the
sample ages, very high fluctuations in density appear where we can
observe some kind of screening effect: more resistant
regions of higher density, protect the looser ones. This model also
gives an insight into a dynamics that exhibits very non-trivial system
size effects. 

Our stochastic model makes an attempt to describe the large strain
behavior of sheared loose granular matter on a mesoscopic level. The
rules of the model do not include any dependence on the total amount
of shear imposed on the medium, nevertheless, a constant friction
angle and slow densification is observed --- a property referred to as
``ageing'' --- which reproduces the experimental results qualitatively
\cite{CompactExp}. By construction, the strain takes place through
local shear bands which initially travel throughout the medium
homogeneously (and hence produce a uniform shear), but which
progressively become more permanent giving rise to more steady shear
bands, a feature also observed experimentally \cite{widthsh}. Our
model reproduces further features seen in experiments and numerical
simulations, including the high frequency fluctuations of the local
shear \cite{Behr}.

In addition, we predict a complex self-organization of these shear
bands, displayed in the inhomogeneities in the local density. This
feature can be studied experimentally, in particular through the use
of X-ray tomography, to access the local density of a sheared medium.
The use of tracer particles could also be helpful in identifying the
inhomogeneous ageing and localization of the shear bands as well as
their sudden changes.

Most of the results presented here for the Euclidean lattice are
closely mirrored by the results on the hierarchical lattice studied in
\cite{TKKRhier}. The recursive topology of this lattice allows a
quantitative analytical understanding of many of the quantities
studied numerically in this paper. This includes elucidating the
mechanism for the breakdown of ergodicity, the slow density evolution,
as well as the behaviour of the of the Hamming distance at late times.

%%%%%%%%%%%%%%%%%%%%%%%%%%%%%%%%%%%%
\section*{Acknowledgments}
%%%%%%%%%%%%%%%%%%%%%%%%%%%%%%%%%%%%

This work was supported by EPSRC, UK, OTKA T029985, T035028.

\end{document}